% Optimal Cloning of Pure States
% Plain TeX, 12 pages, no figures 

\magnification \magstep1

% BLACKBOARD BOLD 
\def\idty{{\leavevmode{\rm 1\ifmmode\mkern -5.4mu\else\kern -.3em\fi I}}}
\def\Ibb #1{ {\rm I\ifmmode\mkern -3.6mu\else\kern -.2em\fi#1}}
\def\Ird{{\hbox{\kern2pt\vbox{\hrule height0pt depth.4pt width5.7pt
    \hbox{\kern-1pt\sevensy\char"36\kern2pt\char"36} \vskip-.2pt
    \hrule height.4pt depth0pt width6pt}}}}
\def\Irs{{\hbox{\kern2pt\vbox{\hrule height0pt depth.34pt width5pt
       \hbox{\kern-1pt\fivesy\char"36\kern1.6pt\char"36} \vskip -.1pt
       \hrule height .34 pt depth 0pt width 5.1 pt}}}}

\def\ibb #1{\leavevmode\hbox{\kern.3em\vrule
     height 1.5ex depth -.1ex width .2pt\kern-.3em\rm#1}}

% THEOREMS  : allow items in proclaim
\def\lessblank{\parskip=5pt \abovedisplayskip=2pt
          \belowdisplayskip=2pt }
\outer\def\iproclaim #1. {\vskip0pt plus50pt \par\noindent
     {\bf #1.\ }\begingroup \interlinepenalty=250\lessblank\sl}

\def\proof#1{\par\noindent {\bf Proof #1}\          % Use as "\proof:"
         \begingroup\lessblank\parindent=0pt}
\def\QED {\hfill\endgroup\break
     \line{\hfill{\vrule height 1.8ex width 1.8ex }\quad}
      \vskip 0pt plus100pt}

\def\Bar{\overline}

\def\bra #1>{\langle #1\rangle}
\def\dim {\mathop{\rm dim}\nolimits}

\def\ket #1{\vert#1\rangle}
\def\ketbra #1#2{{\vert#1\rangle\langle#2\vert}}
\def\norm #1{\left\Vert #1\right\Vert}

  % \Set\Big#1 to force size of \set
% \def\th {\hbox{${}^{{\rm th}}$}\ }  % also in text 
\def\tr {\mathop{\rm tr}\nolimits}
% \def\trace{\mathop{\rm Tr}\nolimits}
% \def\undbar#1{$\underline{\hbox{#1}}$}
% \def\Order{{\bf O}}
% \def\order{{\bf o}}  
% \def\rstr{\hbox{$\vert\mkern-4.8mu\hbox{\rm\`{}}\mkern-3mu$}} 

% LETTERS
\def\phi{\varphi} 
\def\epsilon{\varepsilon} 
\def\3{\ss} 
% \def\Re{\mathchar"023C\mkern-2mu e}
% \def\Im{\mathchar"023D\mkern-2mu m}

%%%%%%%% specific %%%%%%%%%%%%%%%%%%%
\def\H{{\cal H}} 
\def\K{{\cal K}} 
   % redefinition!
\def\FM{{\cal F}} % figure of merit
\def\FMopt{{\widehat{\cal F}}} % best figure of merit
\def\CO{{\widehat T}} %optimal T
\def\dm(#1){{\rm d}\lbrack#1\rbrack}
\def\SU#1{{\rm SU}_{#1}}

\def\quant-ph/9#1#2#3#4#5#6{Report {\tt quant-ph/9#1#2#3#4#5#6}} 
\def\mean#1{\Bar{#1}}
\def\tp#1{^{\otimes#1}}
\def\BC{\gamma} % BlackCow factor  

%%%% CROSS REFERENCE PACKAGE, R.F. Werner, 11-23-1994  
\catcode`@=11
\def\ifundefined#1{\expandafter\ifx\csname
                        \expandafter\eat\string#1\endcsname\relax}
\def\atdef#1{\expandafter\def\csname #1\endcsname}
\def\atedef#1{\expandafter\edef\csname #1\endcsname}
\def\atname#1{\csname #1\endcsname}
\def\ifempty#1{\ifx\@mp#1\@mp} 
\def\ifatundef#1#2#3{\expandafter\ifx\csname#1\endcsname\relax
                                  #2\else#3\fi}
\def\eat#1{}
%%%% CITATIONS %%%%%%%%%%%%%%%%
\newcount\refno \refno=1                
\def\labref #1 #2 #3\par{\atdef{R@#2}{#1}}
\def\lstref #1 #2 #3\par{\atedef{R@#2}{\number\refno}
                              \advance\refno by1}
\def\txtref #1 #2 #3\par{\atdef{R@#2}{\number\refno
      \global\atedef{R@#2}{\number\refno}\global\advance\refno by1}}
\def\doref  #1 #2 #3\par{{\refno=0  
     \vbox {\everyref \item {\reflistitem{\atname{R@#2}}} 
            {\d@more#3\more\@ut\par}\par}}\refskip }
\def\d@more #1\more#2\par
   {{#1\more}\ifx\@ut#2\else\d@more#2\par\fi}
\def\@cite #1,#2\@ver
   {\eachcite{#1}\ifx\@ut#2\else,\hskip0pt\@cite#2\@ver\fi}
\def\cite#1{\citeform{\@cite#1,\@ut\@ver}}
\def\eachcite#1{\ifatundef{R@#1}{{\tt#1??}}{\atname{R@#1}}}
\def\defonereftag#1=#2,{\atdef{R@#1}{#2}}
\def\defreftags#1, {\ifx\relax#1\relax \let\next\relax \else
           \expandafter\defonereftag#1,\let\next\defreftags\fi\next }
\def\@utfirst #1,#2\@ver                                         
   {\author#1,\ifx\@ut#2\afteraut\else\@utsecond#2\@ver\fi}
\def\@utsecond #1,#2\@ver                                         
   {\ifx\@ut#2\andone\author#1,\afterauts\else 
      ,\author#1,\@utmore#2\@ver\fi}
\def\@utmore #1,#2\@ver                                         
   {\ifx\@ut#2\and\author#1,\afterauts\else 
      ,\author#1,\@utmore#2\@ver\fi}
\def\authors#1{\@utfirst#1,\@ut\@ver}                              
\catcode`@=12
%%%%%%%%%%%%%%%%%%%%
\let\REF\labref  
\def\citeform#1{{\bf\lbrack#1\rbrack}}  % appearance \cite{#1}
\let\reflistitem\citeform               % item tag in list
\let\everyref\relax                     % executed before every item
\let\more\relax                         % executed after every subitem
\def\refskip{\vskip 10pt plus 2pt}      % spacing in reflist
\def\colbr{\hskip.3em plus.3em\penalty-100}  % after colon break
\def\combr{\hskip.3em plus4em\penalty-100}   % after comma break
\def\refsecpars{\emergencystretch=50 pt      % done before reflist
                 \hyphenpenalty=100}
%%%%
\def\Bref#1 "#2"#3\more{\authors{#1}:\colbr {\it #2},\combr #3\more} 
\def\Gref#1 "#2"#3\more{\authors{#1}\ifempty{#2}\else:\colbr``#2''\fi
                        ,\combr#3\more} 
\def\Jref#1 "#2"#3\more{\authors{#1}:\colbr``#2'',\combr \Jn#3\more} 
\def\inPr#1 "#2"#3\more{in: \authors{\eds#1}:\colbr
                          ``{\it #2}'',\combr #3\more} 
\def\Jn #1 @#2(#3)#4\more{\hbox{\it#1}\ {\bf#2}(#3)#4\more}
\def\author#1. #2,{\hbox{#1.~#2}}            % single author
%% use as "\sameauthor. {}"   
\def\sameauthor#1{\leavevmode$\underline{\hbox to 25pt{}}$}  
\def\and{, and}   \def\andone{ and}          % separating authors
%% use as    \noinitial. Wolfram Research
\def\noinitial#1{\ignorespaces}
\let\afteraut\relax        
\let\afterauts\relax       
\def\etal{\def\afteraut{, et.al.}\let\afterauts\afteraut
           \let\and,}  
\def\eds{\def\afteraut{(ed.)}\def\afterauts{(eds.)}}
\catcode`@=11

%%%%%%%%%%%%%% EQUATION NUMBERS %%%%%%%%%%%%%%%%%%%%
\newcount\eqNo \eqNo=0 
\def\lasteq{\secNo.\number\eqNo}
\def\deq#1(#2){{\ifempty{#1}\global\advance\eqNo by1
       \edef\n@@{\lasteq}\else\edef\n@@{#1}\fi
       \ifempty{#2}\else\global\atedef{E@#2}{\n@@}\fi\n@@}}
\def\eq#1(#2){\edef\n@@{#1}\ifempty{#2}\else
       \ifatundef{E@#2}{\global\atedef{E@#2}{#1}}%
                       {\edef\n@@{\atname{E@#2}}}\fi
       {\rm($\n@@$)}}
\def\deqno#1(#2){\eqno(\deq#1(#2))} 
\def\deqal#1(#2){(\deq#1(#2))}
\def\eqback#1{{(\advance\eqNo by -#1 \lasteq)}} 
 
\def\eqgroup(#1){{\global\advance\eqNo by1
       \edef\n@@{\lasteq}\global\atedef{E@#1}{\n@@}}}

%%%%%%%%%%%%%%%%% PROCLAIM NUMBERS %%%%%%%%%%%%%%%%%%%%%%%%%
\outer\def\iproclaim#1/#2/#3. {\vskip0pt plus50pt \par\noindent
     {\bf\dpcl#1/#2/ #3.\ }\begingroup \interlinepenalty=250\lessblank\sl}
\newcount\pcNo  \pcNo=0
\def\lastpc{\number\pcNo} % NOT by section
\def\dpcl#1/#2/{\ifempty{#1}\global\advance\pcNo by1
       \edef\n@@{\lastpc}\else\edef\n@@{#1}\fi
       \ifempty{#2}\else\global\atedef{P@#2}{\n@@}\fi\n@@}
\def\pcl#1/#2/{\edef\n@@{#1}%
       \ifempty{#2}\else
       \ifatundef{P@#2}{\global\atedef{P@#2}{#1}}%
                       {\edef\n@@{\atname{P@#2}}}\fi
       \n@@}
\def\Def#1/#2/{Definition~\pcl#1/#2/}
\def\Thm#1/#2/{Theorem~\pcl#1/#2/}
\def\Lem#1/#2/{Lemma~\pcl#1/#2/}
\def\Prp#1/#2/{Proposition~\pcl#1/#2/}
\def\Cor#1/#2/{Corollary~\pcl#1/#2/}
\def\Exa#1/#2/{Example~\pcl#1/#2/} 

%%%%%%%%%%%%%%%% SECTIONS & TABLE OF CONTENTS %%%%%%%%%%%
\font\sectfont=cmbx10 scaled \magstep2            
\def\bgsecti@n #1. #2\e@h{\def\secNo{#1}\eqNo=0}  
\def\bgssecti@n#1. #2\e@h{}                       
\def\secNo{00} 
\def\lookahead#1#2{\vskip\z@ plus#1\penalty-250
  \vskip\z@ plus-#1\bigskip\vskip\parskip
  {#2}\nobreak\smallskip\noindent}
\def\secthead#1. #2\e@h{\vtop{\baselineskip=14pt%
    \sectfont\ifx\n@#1\n@\else\item{#1.}\fi#2}} 
\def\bgsection#1. #2\par{\bgsecti@n#1. #2\e@h
        \lookahead{.3\vsize}{\secthead#1. #2\e@h}} 
\def\bgssection#1. #2\par{\bgssecti@n#1. #2\e@h
        \lookahead{.3\vsize}{\leftline{\bf#1. #2}}} 
\def\bgsections#1. #2\bgssection#3. #4\par{%
        \bgsecti@n#1. #2\e@h\bgssecti@n#3. #4\e@h
        \lookahead{.3\vsize}{\vtop{\secthead#1. #2\e@h\vskip10pt
                             \leftline{\bf#3. #4}}}} 
\def\Acknow#1\par{\ifx\REF\doref
     \bgsection. Acknowledgements\par#1\refsecpars
     \bgsection. References\par\fi}
\catcode`@=12
%%%%%%%%%%%%%%%%%% GENERAL STUFF %%%%%%%%%%%%%%%%%%%%%%%%%%
% for AMS-classification 
\def\class#1 #2*{{#1},} 
\overfullrule=0pt

\let\REF\labref
\REF {BB$^+$} teleport \par
\REF BDEMS Bruss       \par
\REF BEM Bruss2        \par
\REF BH Buzek          \par
\REF DBE Derka         \par
\REF DG Guo            \par
\REF GH GisHut         \par
\REF GM GisMas         \par
\REF GP BlackCow       \par
\REF Hel Helstrom      \par
\REF HB Hillery        \par
\REF Hol Holevo        \par
\REF MP MasPop         \par
\REF Men Anja          \par
\REF vNe vNgame        \par
\REF OP Petz           \par
\REF RS  RSimon        \par
\REF Sti  Stinespring  \par  
\REF Wer CLQ           \par
\REF WW IHQ            \par
\REF WZ WooZu          \par
%%%%%%%%%%%%%%%%%%%%%%%%%

\line{}
\vskip 2.0cm
\font\BF=cmbx10 scaled \magstep 3

{\BF \baselineskip=25pt
\centerline{ Optimal Cloning of Pure States}
}
\vskip 1.0cm
\ifx\draft1\centerline{  Version of \today }\fi
\vskip1cm
 
\centerline{
\bf R.F. Werner
\footnote{$^1$}%
{{\sl 
Inst. f. Mathematische Physik, TU Braunschweig, \hfill\break
Mendelssohnstr.3, 38106 Braunschweig, Germany
}}%
$^,$\footnote{$^{2}$}
{{ \sl Electronic mail:\quad 
   \tt R.Werner@tu-bs.de 
 }}
}
\vskip 1.0cm

{\baselineskip=12pt
\narrower\narrower\noindent
{\bf Abstract.}\
We construct the unique optimal quantum device for turning a finite 
number of $d$-level quantum systems in the same unknown pure state 
$\sigma$ into $M$ systems of the same kind, in an approximation of 
the $M$-fold tensor product of the state $\sigma$. 
\par}
\vskip20 pt

\vfil\eject

%%%%%%%%%%%%%%%%%%%%%%%%
\bgsection 1. Introduction

One of the fundamental features distinguishing quantum theory from 
classical theories is epitomized by the ``No Cloning Theorem'' 
\cite{WooZu}. The ``quantum copiers'' forbidden by this theorem, in 
much the same way as perpetual motion machines are forbidden by the 
Second Law of thermodynamics, are defined as follows: a copier takes 
one quantum system as input and produces as output two systems of 
the same kind. If one now runs experiments in which each input is 
prepared according to the same density matrix, either one of the 
outputs is discarded, and some measurement is then performed on the 
remaining output, one should get the same statistical results as 
measured directly on the inputs, for arbitrary initial preparations 
and final measurements. 

The impossibility of cloning machines is intimately connected to 
other impossible tasks of quantum theory, notably ``joint 
measurement'' and ``teleportation''. It is well-known that there are 
some pairs of quantum observables (such as different spin 
components) which cannot be measured jointly on the same device. 
This statement implies the No-Cloning Theorem, since a quantum 
copier could be operated as a universal joint measuring device: one 
simply applies the two measuring devices in question to the two 
outputs of the copier. Hence a copier is a more powerful machine 
than a joint measuring device. On the other hand, there is a 
hypothetical machine even stronger that the copier: the 
``teleporter'', which is hence also forbidden by the No-Cloning 
Theorem. By definition, ``teleportation'', or ``classical 
teleportation'' to avoid confusion with the fundamental process of 
entanglement enhanced teleportation \cite{teleport}, is the 
transmission of quantum states (or ``quantum information'') on 
classical channels. A teleporting device would consist of a 
measuring apparatus, which produces some classical output (a 
measuring result) from a quantum input, and a reconstruction 
apparatus, which prepares quantum systems, taking the results of the 
previous measurements into account. The criterion for successful 
teleportation is again the impossibility of distinguishing the 
outputs of the overall device from the inputs by statistical 
experiments. To make a copier from a teleporter would be easy: one 
simply has to make copies of the intermediate classical measuring 
results (which is a trivial operation for classical data) and to run 
the reconstruction apparatus on each of these copies. Hence 
classical teleportation is also impossible. 

However, the impossibility of all these devices cannot be the end of 
the story. For example, while the No-Teleportation Theorem declares 
it impossible to determine a quantum state by the classical data 
obtained in a single measurement, it is clearly possible to 
determine quantum states by a run of statistical experiments. In 
fact, according to the statistical interpretation, a quantum state 
is nothing but a mathematical encoding of all data which can be 
collected in this way. Therefore, it must be possible to construct 
devices which take several identically prepared quantum systems as 
an input, make a measurement, and thereby determine the density 
matrix describing the preparation to any desired degree of accuracy. 
This is the problem of quantum state estimation, which has been 
studied by many authors \cite{Holevo,Helstrom,MasPop,Derka}. Of 
course, we can use this classical information to prepare many new 
systems (``clones'') in a state which is a close approximation of 
the input state. Clearly, the quality of the clones will depend on 
the number of initially available input systems. On the other hand, 
there will be no limit to the number of clones obtainable in this 
way, because the classical measuring result can be copied and used 
arbitrarily often. 

More recently, there has been a new twist to this problem coming 
from the observation that if only a given number of clones is 
needed, the procedure via a classical intermediate stage is too 
wasteful \cite{GisMas}. Indeed it has been shown \cite{Hillery} that 
there is, in general, a tradeoff between the number of clones and 
their quality. Clearly, the optimal cloning machine giving a fixed 
number of copies from a fixed number of identically prepared systems 
cannot operate via an intermediate classical stage: It has to stay 
entirely in the quantum world. This paper is is a contribution to 
the theory of such optimal cloning machines. 

There are several variations of the optimal cloning problem, which 
are perhaps best described in the form of a game. Fixed parameters 
in this game are the Hilbert space $\H$ describing the type of 
systems making up the inputs as well as the outputs to the cloning 
device. Its (always finite) dimension will be denoted by $d=\dim\H$. 
Most work so far has been done on the ``qubit case'' $d=2$.
Also fixed will be the number $N$ of input systems and the number 
$M$ of output systems. The game is played by two physicists called 
Alice and Clare. (If the paradigmatical eavesdropper is Eve, why 
should the paradigmatical cloner not be Clare?). Alice's first step 
is to choose a preparation for quantum systems with Hilbert space 
$\H$, as described by a density matrix $\sigma$, say. She then 
proceeds to run her preparing procedure $N$ times, thus producing a 
composite system in the Hilbert space 
$\H\otimes\cdots\otimes\H=\H^{\otimes N}$ (a tensor product of $N$ 
copies) in the state $\sigma^{\otimes N}$, and sends the prepared 
particles to Clare. 
Clare's move is to run a cloning device $T$ of her choice, making 
$M>N$ systems out of the given $N$ systems. (The mathematical 
objects qualifying as ``devices'' in this context will be defined in 
the next section). The next step is to compare Clare's $M$ output 
systems with the state $\sigma^{\otimes M}$, which Alice supplies by 
running her original preparation $M$ times. Clare scores in this 
game, whenever her output is sufficiently similar to $\sigma^{\otimes M}$.
There are different ``figures of merit'' on which Clare's scores 
might be based, resulting in different versions of the cloning game, 
and, possibly, in different ``optimal'' cloning devices $T$. 
Some of the simplest are based on a simple extension of the game: We 
allow Alice a further move, challenging the quality of Clare's 
clones, by choosing some observable. The two then each measure the 
expectation value of this observable on their respective 
$M$-particle states, and the overall score is based on the 
difference of these expectation values. 

Apart from the fine points of the comparison, two basic choices have 
to be made in the rules of this game, leading to four different 
versions of the game. The first choice concerns constraints on the 
initial preparation $\sigma$ done by Alice. For the discussion of 
eavesdropping on quantum cryptography channels it is often of 
interest to allow only a small number of states (e.g., two) 
\cite{GisHut}. Orthogonal states can obviously be cloned perfectly, and 
so can some sets of non-orthogonal ones \cite{Guo}. However, we are 
interested in so-called {\it universal} cloning machines 
\cite{Bruss}, which work on generic (and unknown) inputs. Still 
there is a choice to be made, namely whether or not Alice is 
required to prepare a {\it pure state} $\sigma=\ketbra\phi\phi$ 
given by a wave vector $\phi$. Here the present paper follows most 
of the current literature by imposing purity. The reason is mostly 
that the full mixed-state problem seems to be a lot more involved, 
even in the classical case, and it seems wise to gain a full 
understanding of the simpler case first. 

The second choice to be made in the rules is whether Alice really 
challenges Clare's full $M$-particle output state, or just one clone 
at a time. That is, we could constrain her to challenge Clare's 
result by selecting only one of the $M$ clones, and demanding a one 
particle observable of her choice to be measured on it. This 
constraint on Alice is in keeping with the definition of the quantum 
copier, which also imposes only conditions on one output at a time. 
This ``{\it one particle test}'' version of the cloning problem has 
been considered in the qubit case in several recent papers 
\cite{GisMas,Derka}. It turns out, however, that it is the more 
difficult problem for $d>2$. Therefore, in this paper we will give a 
full analysis of the ``pure state - many particle test'' cloning 
problem for arbitrary $d$. 

The ``pure state - one particle test'' version is settled in the 
$d=2$ case \cite{GisMas}, where the representation theory of $\SU2$ 
makes a full analysis relatively simple. The optimal cloning device 
found by Gisin and Massar is the same as the one found in the 
present paper for the many particle test version. Work on the case 
of general $d$ is under way in Braunschweig with Michael Keyl. The 
optimal cloning devices fit perfectly into the framework for the 
{\it classical limit} (in this case, of $\SU d$-spin systems), set 
up in \cite{CLQ,IHQ,Anja}. In this way, a precise meaning can be 
given to the intuition that cloning very (infinitely) many copies is 
equivalent to the cloning procedure via classical measurement, and 
subsequent preparation. This connection, which requires the 
explanation of more formalism than this paper can take, will be 
explored in a later paper. 

\bgsection 2. Statement of the problem

In order to state the optimal cloning problem precisely we must 
first state what the admissible ``quantum devices'' are among which 
we are looking for an optimal one. There are two ways of approaching 
this problem, which are fortunately equivalent: In either case, each 
device is characterized by its action on quantum states. Thus if the 
input systems are described in a Hilbert space $\H$, and 
consequently the input states are density matrices on $\H$, and the 
output systems are described in a Hilbert space $\K$, a quantum 
device is given by a map $T$ taking density matrices over $\H$ into 
density matrices over $\K$. The first approach to characterizing the 
admissible maps $T$ is the axiomatic one: a minimal requirement for 
$T$ to be consistent with the statistical interpretation of quantum 
theory is that $T$ must respect convex combinations (incoherent 
mixtures) of states. This allows the extension of $T$ to a {\it 
linear} operator from the space of trace class operators over $\H$ 
into the trace class operators over $\K$. This linear operator has 
to take positive elements into positive elements, which is usually 
expressed by calling $T$ a {\it positive} (super-)operator. If this 
condition remains valid if $T$ is applied only to a part of a larger 
system, $T$ is called {\it completely positive}. Since $T$ takes 
density matrices into density matrices, it also has to respect 
normalization (i.e., the trace). Therefore, according to the 
axiomatic definition, an admissible machine must be given by a 
completely positive trace preserving linear operator $T$. The second 
definition of ``admissible devices'' is constructive. It allows only 
operations which can be done by first coupling the given system to 
an auxiliary one (often called the ``ancilla''), then making the two 
interact, as described by a unitary transformation, and finally 
restricting to a suitable subsystem of the combined system by taking 
a partial trace over the ancilla, and perhaps further subsystems. 
Each of these steps is a completely positive trace preserving 
operation, so clearly every quantum device admissible in the 
constructive approach is also admissible by the axiomatic approach. 
But the converse is also true (by virtue of the Stinespring dilation 
theorem \cite{Stinespring}): every linear completely positive trace 
preserving map can be constructed in the way described. 

Let us now turn to the description of figures of merit for quantum 
cloning devices, i.e., on quantitative ways of expressing the 
``closeness'' between the output $T(\sigma^{\otimes N})$ of Clare's 
cloning device and the state $\sigma^{\otimes M}$, which the 
non-existent ideal cloner would achieve. This question has to be 
treated rather carefully for the mixed state versions of the cloning 
game. Possible candidates here are the trace norm difference 
$\norm{T(\sigma^{\otimes N})-\sigma^{\otimes M}}_1$, or perhaps 
another $p$-norm \cite{RSimon} like the Hilbert Schmidt norm, or the 
relative entropy $S(T(\sigma^{\otimes N}),\sigma^{\otimes M})$ 
\cite{Petz}. In principle, the optimal cloner might depend on the 
figure of (de-)merit chosen. However, in the pure state case they 
all lead to the same optimum. In this paper we will use an even 
simpler figure of merit, which makes sense only in the pure case, 
namely the {\it fidelity} 
$\tr\bigl(\sigma\tp M T(\sigma\tp N)\bigr)$, which would be $1$ for 
the non-existent ideal cloner. Good cloning means to bring this 
quantity as close to $1$ as possible, for all input states $\sigma$. 
The worst result 
$$  \FM(T)=\inf_{\sigma,\hbox{\sevenrm pure}} 
    \tr\bigl(\sigma^{\otimes M} T(\sigma^{\otimes N})\Bigr) 
\quad\deqno(FM)$$ 
is taken as the figure of merit. So Clare's and our problem is to 
maximize $\FM(T)$ by a judicious choice of $T$, given $\H$, $N$ and 
$M$. The optimum will be denoted by $\FMopt=\sup_T\FM(T)$, and 
depends on the three integers $d=\dim\H, N$, and $M$.

We note in passing that so far we have only considered the problem 
of minimizing the worst case losses for Clare in a game of the type 
described. It would be interesting to take the game theoretic 
description more seriously, and to ask for the equilibrium points of 
the variants of this game in the sense of von Neumann's theory of 
two-person games \cite{vNgame}. 

\bgsection 3. Description of the optimal cloning machines

In this section we will define the cloning maps, which will be shown 
to be the unique optimal ones in the next section. Since we are 
considering only pure input states $\sigma\tp N$, it suffices to 
consider the action of $T$ on such states and their linear 
combinations. These will be operators on the span of the vectors of 
the form $\phi\otimes\cdots\otimes \phi=\phi\tp N\in\H\tp N$. Our 
first task is to collect some of the basic properties of this space.  

The span of the tensor powers $\phi\tp N$ can be described very 
easily: it is precisely the space of vectors which are invariant 
under all permutations, or the ``Bose subspace'' of $\H^{\otimes N}$ 
in physical terminology. We will denote it by $\H^{\otimes N}_+$. A 
convenient basis in this space is the ``occupation number 
basis'' canonically associated to some basis in the one-particle 
space $\H$. It is labelled by tuples $(n_1,\ldots,n_d)$ with 
$\sum_\kappa n_\kappa=N$. A generating function for this basis is 
given by the tensor power vectors $\phi\tp N$, the variables in the 
generating function being the components $\phi_1,\ldots,\phi_d$ of 
$\phi$ in the given basis of $\H$. Explicitly, 
$$ \phi^{\otimes N}=\sqrt{N!}\sum_{n_1,\ldots,n_d}
                   \prod_{\kappa=1}^d{\phi_\kappa^{n_\kappa}\over
                                         \sqrt{n_\kappa!}}
                    \ket{n_1,\ldots,n_d}
\quad.\deqno(genFunc)$$
It is easily checked, using this basis that the dimension of 
$\H\tp N$ is 
$$ \dm(N)=(-1)^N{-d\choose N}={d+N-1\choose N} 
\quad,\deqno(dN)$$ 
where $d=\dim\H$. 
We will denote by $s_N$ the orthogonal projection of 
$\H\tp N$ onto $\H\tp N_+$. A crucial feature of the 
symmetric subspace is that the unitary operators $U\tp N$ leave it 
invariant, and act on it irreducibly. That is to say, any operator 
$A$ supported by $\H\tp N_+$ ($A=As_N=s_NA$), which commutes with 
all $U\tp N$, must be a multiple of $s_N$, i.e., of the identity 
operator on $\H\tp N_+$.

The optimal cloning map has to take density operators on 
$\H\tp N_+$ to operators on $\H\tp M$. An easy way to 
achieve such a transformation is to tensor the given operator $\rho$ 
with the identity operators belonging to tensor factors $(N+1)$ 
through $M$, i.e., to take 
$\rho\mapsto\rho\otimes\idty\tp{(M-N)}$. 
This breaks the symmetry between the clones, making $N$ perfect 
copies and $(N-M)$ states, which are worst possible ``copies''. 
Moreover, it does not 
map to states on the Bose sector $\H\tp M_+$, which would 
certainly be desirable, as the target states $\sigma\tp M$ are 
supported by that subspace. An easy way to remedy both 
defects is to compress the operator to the symmetric subspace with 
the projection $s_M$. With the appropriate normalization this is our 
definition of the cloning map, later shown to be optimal: 
$$ \CO(\rho)= {\dm(N)\over \dm(M)}\ 
            s_M\bigl(\rho\otimes\idty^{\otimes(M-N)}\bigr)s_M
\quad.\deqno(optclo)$$
Complete positivity is obvious from the form of $\CO$. So in order to 
verify that this is a legitimate cloning map, we only have 
to check that the normalization factor is chosen correctly to make 
$\CO$ trace-preserving. To begin with, $\tr \CO(\rho)$ is a linear 
functional of $\rho$, and can hence be written as $\tr(\rho X)$, for 
a suitable positive operator $X$ on $\H\tp N_+$. From the {\it 
covariance} of $\CO$, i.e., the property 
$$  T\Bigl(U\tp N\rho U^{*\otimes N} \Bigr)
   =U\tp M T(\rho)U^{*\otimes M} 
\quad,\deqno(cov)$$
one concludes that $X$ commutes with $U\tp N$ and, by irreducibility, 
$X$ must be a multiple of the identity. It remains to be shown that 
this multiple is $1$ or, equivalently, that the trace of {\it some} 
density matrix is preserved by $\CO$. To this end we consider the 
maximally mixed density matrix $\tau_N=\dm(N)^{-1}s_N$ on $\H\tp N_+$, 
which is also characterized as the unique density matrix on 
$\H\tp N_+$ invariant under sitewise rotations 
$\rho\mapsto U\tp N\rho U^{*\otimes N}$. Then 
$\CO(\tau_N)=\dm(M)^{-1}s_M(s_N\otimes\idty^{M-N})s_M
    =\dm(M)^{-1}s_M=\tau_M$.
Hence $\CO$ as defined in \eq(optclo) is trace preserving. 

The value of $\FM(\CO)$ is determined by observing that, for a pure 
state $\sigma$ on $\H$, $\sigma\tp M$ is a one-dimensional 
projection, which is smaller than both $s_M$ and 
$(\sigma\tp N\otimes\idty\tp{(M-N)})$. Hence 
$$ \FM(\CO)={\dm(N)\over \dm(M)}\tr\bigl(\sigma\tp M s_M
           (\sigma\tp N\otimes\idty\tp{(M-N)})s_M\bigr)
         ={\dm(N)\over \dm(M)}\tr\bigl(\sigma\tp M\bigr)
         ={\dm(N)\over \dm(M)}
\quad.\deqno(FMopt)$$

We conclude this section by computing the performance of $\CO$ with 
respect to the one-particle-test version of the cloning problem. 
Some of our considerations will be valid for any cloning map $T$ 
(not necessarily $T=\CO$), which maps density matrices on 
$\H\tp N_+$ into density matrices on $\H\tp M_+$, and satisfies the 
covariance condition \eq(cov). For any density matrix $\rho$ on 
$\H\tp N$, we denote by $R(\rho)$ its one-site restriction defined 
by $\tr(R(\rho)A)=\tr(\rho(A\otimes\idty\tp{(N-1)}))$. Consider the 
one-site restriction  $R(T(\sigma\tp N))$. By covariance of $T$, 
this must be a density matrix on the one-site Hilbert space $\H$, 
commuting with all unitaries $U$, which commute with $\sigma$. Hence 
we can write it as 
$$ R(T(\sigma\tp N))
     =\BC(T)\sigma +(1-\BC(T))\tau_1
\quad,\deqno(BC)$$
where $\tau_1=d^{-1}\idty$ is the totally mixed density matrix on 
$\H$. By covariance of $T$, the number $\BC(T)$ does not depend on 
$\sigma$, and is called the {\it Black Cow factor$\,^{*}$} of $T$.
\vfootnote{*}{ 
    The reason for this terminology is that it plays an important 
    role in discussions of the cloning problem started by Chiara 
    Machiavello and Artur Ekert at the Black Cow Caf\'e in 
    Croton-on-Hudson, NY, and further clarified in collaboration 
    with Dagmar Bru\ss\ \cite{Bruss2}. I learned about this line of 
    argument from a set of ``Black Cow Notes'' by Nicolas Gisin and 
    Sandu Popescu.} 
Surprisingly, it is useful also for the discussion of ``cloning 
in stages'' from $N$ to $M$ to $R$ systems, even though in the 
second stage the cloner from $M$ to $R$ systems no longer finds a 
product density matrix  $\sigma\tp M$. In fact, on the right hand 
side of \eq(BC) we can write $R(\sigma\tp N)$ for $\sigma$, and it 
is clear that \eq(BC) becomes 
$$ R(T(\rho))
     =\BC(T)R(\rho)+(1-\BC(T))\tr(\rho)\tau_1
\quad,\deqno(BCro)$$
for all $\rho$ in the linear span of the operators $\sigma\tp N$. 
But these are {\it all} density matrices on $\H\tp N_+$: this can be 
seen by inserting the expansion \eq(genFunc) into 
$\sigma\tp N=\ketbra{\phi\tp N}{\phi\tp N}$, and observing that from 
the resulting power series in $\phi_\kappa$ and $\Bar{\phi_\kappa}$ 
the coefficients $\ketbra nm$ can be extracted. Hence \eq(BCro) 
holds for all cloning maps $T$ satisfying the assumptions stated at 
the beginning of this paragraph. 

As a corollary we obtain the equation 
$\BC(T_{RM}T_{MN})=\BC(T_{RM})\BC(T_{MN})$, for cloning in stages. 
Since the family of optimal cloners defined by \eq(optclo) obviously 
satisfies the concatenation property $\CO_{RM}\CO_{MN}=\CO_{RN}$, we 
find that the Black Cow Factor for these must be of the form 
$\BC(\CO_{MN})=\gamma_N/\gamma_M$. 

To compute $\BC(\CO_{MN})$ for \eq(optclo), we use the normalization property of 
$\CO$ in the form 
$\tr(s_M\sigma\tp N\otimes\idty\tp{(M-N)})=\dm(M)/\dm(N)$, for any 
pure $\sigma$. 
Then, on the one hand,  we find that 
$$\tr\bigl(\sigma R\CO(\sigma\tp N)\bigr)
    = \BC(\CO)+(1-\BC(\CO))/d
\quad,$$
and, on the other hand,
$$\eqalign{ \tr\bigl(\sigma R\CO(\sigma\tp N)\bigr)
    &= \tr\bigl((\sigma\otimes\idty\tp{(M-1)}) 
              \CO(\sigma\tp N)\bigr) \cr
    &= {1\over M}\sum_k \tr\bigl(\sigma^{(k)}
              \CO(\sigma\tp N)\bigr) \cr
    &= {\dm(N)\over M\ \dm(M)}\sum_k \tr\bigl(\sigma^{(k)} s_M
              (\sigma\tp N\otimes\idty\tp{(M-N)})s_M\bigr) \cr
    &= {\dm(N)\over M\ \dm(M)}\sum_k \tr\bigl(\sigma^{(k)}
              (\sigma\tp N\otimes\idty\tp{(M-N)})s_M\bigr) \cr
    &= {\dm(N)\over M\ \dm(M)}\left\lbrace 
            N {\dm(M)\over \dm(N)}+ (M-N){\dm(M)\over \dm(N+1)}
                          \right\rbrace     \cr
    &={N\over M} +{M-N\over M} {N+1\over d+N}  
\quad,\cr}$$
where in the second line we used the abbreviation $\sigma^{(k)}$ for 
the tensor product of $M$ operators, all of which are $\idty$, 
except the $k^{\rm th}$, which is $\sigma$. At the fourth equality we 
used that the sum $\sum_k\sigma^{(k)}$ commutes with permutations and 
hence with $s_M$.
Solving for $\BC(\CO)$ we find the Black Cow Factor of \eq(optclo)
to be 
$$ \BC(\CO)={N\over d+N}\, {d+M\over M}
\quad.\deqno(BCopt)$$
This is a quotient, as expected.  Specializing to $d=2$ we find this 
result also in agreement with the value found in \cite{Bruss2} by
combining the Black Cow concatenation argument with the previously 
determined optimal value for state determinations. Again this is to 
be expected, because the optimal cloner found in the 
one-particle-test version of the problem (for $d=2$) agrees with 
\eq(optclo). We conjecture that the equality of the optimal 
solutions to the one-particle-test and many-particle-test versions 
of the cloning problem coincide also for $d>2$. 

\bgsection 4. Proof of optimality

In this section we will prove the optimality of the cloning map 
$\CO$, defined in \eq(optclo), with respect to the figure of merit 
$\FM$ from \eq(FM). Let 
$$  \FMopt=\sup_T\FM(T) 
\quad\deqno(supFM)$$
be the best bound for $\FM(T)$. Since $\FM$ is an infimum of 
continuous functions, it is an upper semicontinuous function, and 
since the set of admissible $T$ is compact (bounded and closed in a 
finite dimensional vector space), the supremum \eq(supFM) is 
attained, i.e., optimal cloners with $\FM(T)=\FMopt$ do exist. 

For a pure state $\sigma$, rotated by unitary $U$ on $\H$, we will 
write $\sigma_U\equiv U\sigma U^*$. The average of any cloning map 
with respect to sitewise rotations will be denoted by 
$$ \mean T(\rho)=\int\!\!\!dU\ 
           U^{*\otimes M}T\Bigl(U^{\otimes N}\rho
           U^{*\otimes N} \Bigr)U^{\otimes M}
\quad,\deqno(mean)$$
where ``$dU$'' denotes the integration with respect to the 
normalized Haar measure of the unitary group of $\H$. Then $\mean T$ 
is again an admissible cloning map, and $T=\mean T$ if and only if 
$T$ satisfies the covariance condition \eq(cov).

\proclaim Theorem. 
For any cloning map from $N$ to $M$ systems, 
$$\FM(T)\leq \dm(N)/\dm(M)
\quad,$$
with equality if and only if $T=\CO$. 

\proof:
Let $T$ be an optimal cloning device, i.e., $\FM(T)=\FMopt$.
Then, for every pure $\sigma$, we have 
$$\tr\bigl(\sigma\tp M \mean T(\sigma\tp N)\bigr)
       = \int\!\!\!dU\ \tr\bigl(\sigma_U\tp M T(\sigma_U\tp N)\bigr)
       \geq \int\!\!\!dU\ \FM(T)  =\FMopt
\quad.$$
Since the left hand side is independent of $\sigma$, it is also 
equal to $\FM(\mean T)$, hence $\FM(\mean T)\geq\FMopt$. By 
definition of $\FMopt$ we also have $\FMopt\geq\FM(\mean T)$, i.e.,  
$\FM(\mean T)=\FMopt$. Hence the integral over the positive 
quantities 
$\FMopt- \tr\bigl(\sigma_U\tp M T(\sigma_U\tp N)\bigr)$ vanishes, 
which implies that 
$$ \tr\bigl(\sigma\tp M T(\sigma\tp N)\bigr)
    =\FM(T)=\FMopt
$$
for all $\sigma$. 

Next consider the rotation invariant density matrix 
$\tau_N=\dm(N)^{-1}s_N$ on 
the symmetric subspace. Since $\mean T$ commutes with rotations, 
$\tau_N$ has to be mapped into a likewise rotation invariant density 
matrix on $\H\tp M$. In particular, because the representation 
$U\tp M$ restricted to the symmetric subspace is irreducible, we 
must have 
$$ \mean T\Bigl({s_N\over \dm(N)}\Bigr)
      =\lambda{s_M\over \dm(M)} +(1-\lambda) \hbox{Rest}
\quad,$$
where ``Rest'' stands for a density matrix orthogonal to $s_M$, and 
$0\leq\lambda\leq1$. 
We now use that $\mean T(s_N-\sigma\tp N)$ must be a positive 
operator. Taking its trace with $\sigma\tp M$ we thus find that 
$$  0\leq \tr\Bigl(\sigma\tp M \mean T(s_N-\sigma\tp N)\Bigr) 
     =\lambda {\dm(N)\over \dm(M)} - \FMopt
\quad.\deqno*()$$
Hence $\FMopt\leq\lambda{\dm(N)/\dm(M)}\leq{\dm(N)/\dm(M)}$.
Since we have already seen in \eq(FMopt) that 
$\FM(\CO)=\dm(N)/\dm(M)$, we have shown that $\FMopt$ is equal to 
this value, and $\CO$ is indeed optimal. 

It remains to be shown that $\CO$ is the only cloning map 
achieving this value. 
From the last string of inequalities we see that for any optimal 
cloner we must have $\lambda=1$. This is equivalent to saying that 
$\mean T(\sigma\tp N)$ is supported by the symmetric subspace for 
all $\sigma$ and, since $\mean T$ is an integral over rotated copies 
of $T$, the same conclusion also holds for $T$.
Moreover,  for an optimal cloner $T$, the right hand side of \eq*() 
has to vanish. This is again an integral with respect to $dU$ over 
a positive function, which hence has to vanish, too:  
$$\tr\bigl(\sigma\tp M T(s_N-\sigma\tp N)\bigr)=0
\quad.\deqno**()$$ 
Since the second term in this expression 
was already shown to be equal to $\FMopt$ for all $\sigma$, we 
conclude that $\tr\bigl(\sigma\tp M T(s_N)\bigr)=\FMopt$ for all 
$\sigma$. The operators $\sigma\tp M$ span the space of operators on 
$\H\tp M_+$. Hence this equation is equivalent to 
$T(s_N)=\FMopt s_M$.

To further exploit the optimality condition, we introduce the 
Stinespring dilation \cite{Stinespring}of $T$ in the form
$$ T(\rho)=\FMopt\ V^*(\rho\otimes\idty_\K)V
\quad,$$
where $V:\H\tp M_+\to\H\tp N_+\otimes\K$ for some auxiliary Hilbert 
space $\K$, and $\rho$ is an arbitrary density matrix on $\H\tp N_+$. 
We have included the factor $\FMopt$ in this definition, so that for 
an optimal cloner $V^*V=\idty$. The optimality condition\eq**(), 
written in terms of $V$ becomes 
$$  \bra\phi\tp M, V^*((s_N-\sigma\tp N)\otimes\idty_\K)V\phi\tp M>
    =\norm{((s_N-\sigma\tp N)\otimes\idty_\K)V\phi\tp M}^2
    =0
\quad,$$ 
where $\sigma$ is the one-dimensional projection to $\phi\in\H$. 
Equivalently, 
$((s_N-\sigma\tp N)\otimes\idty_\K)V\phi\tp M=0$, which is to say 
that $V\phi\tp M$ must be in the subspace $\phi\tp N\otimes\K$ for 
every $\phi$. 

So we can write $V\phi\tp M=\phi\tp N\otimes\xi(\phi)$, with 
$\xi(\phi)\in\K$ some vector depending in a generally non-linear way 
on the unit vector $\phi\in\H$. From the above observation that $V$ 
must be an isometry we can calculate the scalar products of all the 
vectors $\xi(\phi)$: 
$$\eqalign{ \bra\phi,\psi>^M
    & = \bra\phi\tp M,\psi\tp M> 
      = \bra V\phi\tp M,V\psi\tp M>    \cr 
     &= \bra \phi\tp N\otimes\xi(\phi),\psi\tp N\otimes\xi(\psi)> 
      = \bra\phi,\psi>^N\ 
         \bra\xi(\phi),\xi(\psi)>_\K \quad,\cr\hbox{i.e., }\qquad 
            \bra\xi(\phi),\xi(\psi)>_\K &=\bra\phi,\psi>^{M-N} 
      =\bra\phi\tp{M-N},\psi\tp{M-N}> 
\quad.}$$ 
This information is 
sufficient to compute all matrix elements 
$\bra\psi_1\tp M, T\bigl(\ketbra{\phi_1\tp N}% 
     {\phi_1\tp N}\bigr)\psi_2\tp M>$, i.e., 
$T$ is uniquely determined and equal to $\CO$. 
\QED

\let\REF\doref
\Acknow
This paper is a response to the many discussions about the cloning 
problem I had with participants of the ISI Workshop on Quantum 
Computing in Torino in July 1997, notably N.~Gisin, C.~Machiavello, 
S.~Massar, A.~Ekert, D.~Bru\ss, S.~Popescu, and W.~van~Dam. 
I would like to thank several of these, and also P.~Zanardi for 
comments on an earlier version, and M.~Keyl for a critical reading 
of the manuscript.

\REF {BB$^+$} teleport \Jref
    C.H. Bennett, G. Brassard, C. Crepeau, R. Jozsa, A. Peres, 
    W.K. Wootters
    "Teleporting an unknown quantum state via dual classical and 
    Einstein-Podolsky-Rosen channels"
    Phys.Rev.Lett. @70(1993) 1895--1899

\REF BDEMS Bruss \Gref
    D. Bru\ss, D.P. DiVincenzo, A. Ekert, C. Machiavello, 
    J.A. Smolin
    "Optimal universal and state-dependent cloning"
    \quant-ph/9705038
    
\REF BEM Bruss2 \Gref
    D. Bru\ss, A. Ekert, C. Machiavello
    "Optimal universal cloning and state estimation"
    \quant-ph/9712019

\REF BH Buzek \Gref
    V. Bu{\v z}ek, M. Hillery
    "Quantum copying: beyond the no-cloning theorem"
    \quant-ph/9607018

\REF DBE Derka \Gref
    R. Derka, V. Bu{\v z}ek, A.K. Ekert
    "Universal algorithm for optimal estimation of quantum states 
    from finite ensembles"
    \quant-ph/9707028

\REF DG Guo \Gref 
    L.-M. Duan, G.-C. Guo
    "Two non-orthogonal states can be cloned by a unitary-reduction 
    process"
    \quant-ph/9704020

\REF GH GisHut \Jref
    N. Gisin, B. Huttner
    "Quantum cloning, eavesdropping and Bell's inequality"
    Phys.Lett. A @228(1997) 13

\REF GM GisMas \Gref
    N. Gisin, S. Massar
    "Optimal quantum cloning machines"
    \quant-ph/9705046

\REF GP BlackCow \Gref
    N. Gisin, S. Popescu
    "The Black Cow notes" 
    unpublished

\REF Hel Helstrom \Bref
    C.W. Helstrom
    "Quantum detection and estimation theory"
    Academic Press, New York 1976

\REF HB Hillery \Jref
    M. Hillery, V. Bu{\v z}ek
    "Quantum copying: Fundamental inequalities"
    Phys.Rev. A @56(1997) 1212--1216

\REF Hol Holevo \Bref
    A.S. Holevo
    "Probabilistic and statistical aspects of quantum theory"
    North-Holland, Amsterdam 1982

\REF MP MasPop \Jref
    S. Massar, S. Popescu
    "Optimal extraction of information from finite quantum ensembles"
    Phys.Rev.Lett. @74(1995) 1259--1263

\REF Men Anja \Gref
    A. Menkhaus
    "Der klassische Limes von Spinsystemen"
    Diplomarbeit, Osnabr\"uck 1996
    Internal report,
     {\tt \hfill\break
ftp://ftp.physik.uni-osnabrueck.de/pub/werner/ClassLimit/anja.ps}

\REF vNe vNgame \Jref
    J. von Neumann
    "Zur Theorie der Gesellschaftsspiele"
    Math.Ann @100(1928) 295--320

\REF OP Petz \Bref
    M. Ohya, D. Petz 
    "Quantum entropy and its use" 
    Springer Verlag, Heidelberg 1993.

\REF RS  RSimon \Bref
    M. Reed, B. Simon
    "Methods of Modern Mathematical Physics, Vol. II"
    (Appendix on abstract interpolation)
    Academic Press, New York 1975

\REF Sti  Stinespring    \Jref
     W.F. Stinespring
     "Positive functions on C*-algebras"
     Proc.Amer.Math.Soc. @6(1955)211--216

\REF Wer CLQ \Gref
    R.F. Werner
    "The classical limit of quantum theory" 
    \quant-ph/9504016

\REF WW IHQ \Jref
    R.F. Werner, M.P.H. Wolff
    "Classical mechanics as quantum mechanics with infinitesimal 
    $\hbar$"
    Phys.Lett.A @202(1995) 155--159

\REF WZ WooZu \Jref
    W.K. Wootters, W.H. Zurek
    "A single quantum cannot be cloned"
    Nature @299(1982) 802--803

\end